# Analysis and Synthesis of Disturbance Observer-based Digital Robust Motion Control Systems in State Space

Emre Sariyildiz, *Senior Member, IEEE*

*Abstract*— Despite its extensive applications in motion control, there remains a lack of systematic analysis and synthesis methods capable of ensuring high- stability and performance for Disturbance Observer (DOb)-based robust motion controllers. The development of such methods is essential for achieving precise disturbance rejection, enhanced robustness, and high-performance motion control. In response to this need, this paper proposes a novel analysis and synthesis method for DOb-based digital robust motion controllers. By employing a unified state-space design framework, the proposed synthesis approach facilitates the implementation of both conventional zero-order (ZO) and high-order (HO) DObs, offering a systematic design method applicable to a wide range of motion control systems. Furthermore, this design method supports the development of advanced DObs (e.g., the proposed High-Performance (HP) DOb in this paper), enabling more accurate disturbance estimation and, consequently, enhancing the robust stability and performance of motion control systems. Lyapunov's direct method is employed in the discrete-time domain to analyse the stability of the proposed digital robust motion controllers. The analysis demonstrates that the proposed DObs are stable in the sense that the estimation error is uniformly ultimately bounded when subjected to bounded disturbances. Additionally, they are proven to be asymptotically stable under specific disturbance conditions, such as constant disturbances for the ZO and HP DObs. Stability constraints on the design parameters of the DObs are analytically derived, providing effective synthesis tools for the implementation of the digital robust motion controllers. The discrete-time analysis facilitates the derivation of more practical design constraints. The proposed analysis and synthesis methods have been rigorously validated through experimental evaluations, confirming their effectiveness.

## I. INTRODUCTION

IN THE last three decades, DOb has been one of the most widely used robust control techniques in the literature, owing to its intuitive design approach, ease of implementation due to its low computational load, and its ability to deliver high-performance in the presence of significant disturbances [1–3]. It has been applied to many motion control problems across various fields, spanning from the precise positioning of a hard-disk drive [4] and the interaction force estimation of an exoskeleton robot [5] to compensating constant and varying time-delays in network systems [6]. These high-performance engineering applications have generally been conducted using basic intuitive design approaches in the frequency domain due to its simplicity [7–10]. Despite reported successful engineering applications, the stability and performance of the robust motion controllers are highly contingent upon the designers' expertise when basic intuitive design approaches are employed in DOb synthesis. This results in poor stability and performance in many robust motion control applications [1, 11]. To effectively resolve this issue, it is crucial to establish systematic analysis and synthesis tools that ensure robust stability and high-performance for DOb-based robust motion control systems.

In general, a DOb can be implemented using two primary observer design approaches. The first approach is the minimum-order observer-based disturbance estimation method, known as DOb-based control, proposed by Ohnishi in 1983 [12]. The second approach is Han's full-state observer-based disturbance estimation method, commonly referred to as Active Disturbance Rejection Control (ADRC), which was introduced in the 1990s [13]. Both observers can be implemented using linear or nonlinear design techniques. Despite the reported advantages of nonlinear DObs in the literature, linear DObs are commonly employed in motion control applications due to their simpler analysis and synthesis procedures. [1, 13–15].

Traditionally, frequency-domain analysis and synthesis methods have been employed for the development of DOb-based robust motion controllers, predominantly within the continuous-time framework [1, 2, 11–17]. However, this approach presents two significant limitations. First, frequency-domain methods are typically effective only for relatively simple motion control systems, as they fail to adequately represent more complex system dynamics in a straightforward manner [2, 18]. This results in intricate transfer functions in many DOb-based robust motion control applications, such as those employing high-order DObs [7, 19] and the robust motion control of series elastic actuators [20]. Second, continuous-time analysis methods are insufficient for fully capturing certain dynamic behaviours of DOb-based digital robust motion controllers, leading to unexpected stability and performance problems, as reported in [11, 16, 17]. For instance, digital robust motion controllers may exhibit degraded stability when the nominal inertia and bandwidth of the DOb exceed specific design thresholds, in contrast to continuous-time analysis methods, which predict only enhanced robust stability as these parameters are increased [21, 22].

To elucidate these unexpected dynamic behaviours and ensure high-performance motion control, numerous researchers have analysed and synthesised DOb-based digital robust motion controllers within the discrete-time domain [11, 23, 24]. While bilinear transformation method was employed to implement the digital robust motion controllers in [25], the design parameters of digital DObs were tuned using different loop-shaping control techniques, e.g., sensitivity optimisation [26], optimal plant shaping [27], and maximising robustness margin [28]. Different discretisation methods were evaluated in [29, 30], highlighting their substantial impact on system stability and performance. A rigorous stability analysis elucidating the unexpected dynamic behaviours of DOb-based digital robust motion controllers was recently proposed using precise discrete transfer functions and Bode plots in [11, 22]. Although frequency-domain analysis and synthesis methods are useful for certain DOb-based robust motion controllers, they typically necessitate complex transfer functions, which can obscure the understanding of system dynamics and impede the development of effective controllers in many robust motion control applications [22]. This not only impairs the performance of DOb-based robust motion controllers but also significantly constraints their broader applicability. Therefore, it is essential to develop new analysis and synthesis methods for DOb-based digital robust motion control systems.

In this paper, a unified analysis and synthesis method is proposed for DOb-based digital robust motion control systems in state-space. The proposed method not only facilitates the synthesis of the conventional ZO and HO DObs, but also enables the synthesis of advanced DOb configurations. One of the primary advantages of the proposed synthesis method is that it maintains a straightforward design process for the digital DOb, regardless of the observer's order. Additionally, the development of an advanced DOb enhances the performance of the robust motion controller by providing more accurate disturbance estimation compared to conventional DObs. The stability of the proposed DObs and digital robust motion controllers is analysed using Lyapunov's direct method within a unified framework. It is demonstrated that the proposed digital robust motion controllers maintain uniform ultimate boundedness under bounded disturbances and achieve asymptotic stability under certain stringent disturbance conditions, provided that the design parameters of the digital DOb conform to the stability constraints outlined in the paper. To the best of our knowledge, this paper, for the first time, analytically derives the design constraints of digital DObs using a rigorous stability analysis in state-space. This can pave the way for the development of high-performance DOb-based digital robust motion control systems in various engineering applications. Experimental results are presented to validate the effectiveness of the proposed analysis and synthesis methods.

Within this context, the chief contributions of this paper are i) proposing a unified analysis and synthesis method for DOb-based digital robust motion control systems within the state-space framework, ii) introducing a novel high-performance DOb that can significantly enhance disturbance estimation accuracy, thereby improving the performance of robust motion control applications, and iii) deriving the stability constraints of digital DObs analytically, and providing effective design tools for implementing DOb-based digital robust motion controllers.

The rest of the paper is organised as follows. In Section II, a novel synthesis method is proposed for DOb-based digital robust motion control systems in state-space. In Sections III and IV, a new stability analysis is presented for the digital DOb and robust motion controllers, and the design constraints of DObs are derived analytically. In Section V, experimental results are presented to validate the proposed analysis and synthesis methods. The paper is concluded in section VI.

## II. DOb Synthesis in the Discrete-time Domain

### A. Servo System Model:

The continuous-time uncertain and nominal dynamic models of a servo system can be represented in state space as follows:

$$\begin{aligned} Uncertain: \dot{\mathbf{x}}(t) &= \mathbf{A_C}\mathbf{x}(t) + \mathbf{B_C}u(t) - \mathbf{D_C}\tau_d(t) \\ Nominal: \dot{\mathbf{x}}(t) &= \mathbf{A_{Cn}}\mathbf{x}(t) + \mathbf{B_{Cn}}u(t) - \mathbf{D_{Cn}}\tau_{dn}(t) \end{aligned} \quad (1)$$

where $\mathbf{A_{C*}} = \begin{bmatrix} 0 & 1 \\ 0 & b_{m*}/J_{m*} \end{bmatrix}$, and $\mathbf{B_{C*}} = \mathbf{D_{C*}} = \begin{bmatrix} 0 \\ 1/J_{m*} \end{bmatrix}$ are the state matrix, and control and disturbance input vectors in which $J_{m*}$ and $b_{m*}$ represent the inertia and viscous friction coefficient, respectively; $u(t)$ is the control input; $\tau_d(t)$ and $\tau_{dn}(t)$ are the external and fictitious nominal disturbance variables, respectively; $\mathbf{x}(t)$ is the state vector, comprising the position and velocity states of the servo system; and $*$ is blank and $n$ for the uncertain and nominal dynamic models, respectively. The reader is referred to [1] for a more detailed dynamic model of a servo system affected by internal and external disturbances.

Equation (2) represents the uncertain and nominal dynamic models of a servo system in the discrete-time domain when the motion controller is implemented using a zero-order hold.

$$\begin{aligned} Uncertain: \mathbf{x}((k+1)T_s) &= \mathbf{A_D}\mathbf{x}(kT_s) + \mathbf{B_D}u(kT_s) - \mathbf{\Pi_D}(kT_s) \\ Nominal: \mathbf{x}((k+1)T_s) &= \mathbf{A_{Dn}}\mathbf{x}(kT_s) + \mathbf{B_{Dn}}u(kT_s) - \mathbf{\Pi_{Dn}}(kT_s) \end{aligned} \quad (2)$$

where $\mathbf{A_{D*}} = e^{\mathbf{A_{C*}}T_s}, \mathbf{B_{D*}} = \int_0^{T_s} e^{\mathbf{A_{C*}}\lambda}\mathbf{B_{C*}}d\lambda,; T_s$ is the sampling time; $\mathbf{\Pi_{D*}}(kT_s) = \int_0^{T_s} e^{\mathbf{A_{C*}}\lambda}\mathbf{D_{C*}}\tau_{d*}((k+1)T_s - \lambda)d\lambda$ is the discrete disturbance vector; and $*$ is similarly blank and $n$ in the discrete uncertain and nominal servo system models, respectively. It is noted that hereafter, the discrete-time variables at $kT_s$ seconds are represented using $\bullet(k)$ rather than $\bullet(kT_s)$ for the sake of brevity.

### B. DOb Synthesis:

To estimate disturbances exerted on a servo system, an approximate disturbance model must be incorporated in the synthesis of a DOb [1]. This model is essential for accurately capturing the characteristics of disturbances, which in turn plays a critical role in enhancing the disturbance estimation accuracy and, consequently, improving the overall motion control performance.

Let us introduce a general approximate disturbance model using the $m^{th}$-order Taylor series expansion of the nominal disturbance vector, as expressed in Eq. (3).

$$\mathbf{\Pi_{Dn}}(k) = \int_0^{T_s} e^{\mathbf{A_{Cn}}\lambda} \mathbf{D_{Cn}} \left( \sum_{i=0}^{m} \left( \frac{1}{i!} \overset{(i)}{\tau}_{dn}(k)(T_s - \lambda)^i \right) + R_m(k) \right) d\lambda \quad (3)$$

where $R_m(k) = \frac{1}{(m+1)!} \overset{(m+1)}{\tau}_{dn}(\xi_k)(T_s - \lambda)^{m+1}$ is the truncation error of order $m+1$, i.e., $R_m(k) = O(m+1)$, in which $\overset{(i)}{\tau}_{dn}$ is the $i^{th}$ order derivative of $\tau_{dn}(k)$, and $kT_s \leq \xi_k \leq (k+1)T_s$ is an uncertain time within the sampling period.

To synthesise the $m^{th}$-order DOb that can estimate the nominal disturbance variable and its derivatives up to order $m$, let us employ the auxiliary variables given by:

$$\begin{aligned} z_0(k) &= \tau_{dn}(k) + \mathbf{L_0}^T \mathbf{x}(k) \\ &\vdots \\ z_m(k) &= \overset{(m)}{\tau}_{dn}(k) + \mathbf{L_m}^T \mathbf{x}(k) \end{aligned} \quad (4)$$

where $\mathbf{L_0}, \mathbf{L_1}, \cdots$ and $\mathbf{L_m} \in R^2$ are the observer gain vectors, which are tuned in Section III.

The dynamics of the auxiliary variables can be derived by substituting Eqs. (2) and (3) into Eq. (4) as follows:

$$\mathbf{z}(k+1) = \mathbf{\Gamma z}(k) + \mathbf{\Omega_x x}(k) + \mathbf{\Omega_u} u(k) + \mathbf{\Lambda}(k) \quad (5)$$

where $\mathbf{z}(k) = \begin{bmatrix} z_0(k) & \cdots & z_m(k) \end{bmatrix}^T \in R^m$ is the auxiliary variable vector, $\mathbf{\Gamma} \in R^{m \times m}$ is the auxiliary variable state matrix, $\mathbf{\Omega_x} \in R^{m \times 2}$ is the coefficient matrix of $\mathbf{x}(k)$, $\mathbf{\Omega_u} \in R^m$ is the control input vector, and $\mathbf{\Lambda}(k) = \begin{bmatrix} 0,0 & \cdots & \overset{(m)}{\tau}_{dn}(k+1) - \overset{(m)}{\tau}_{dn}(k) \end{bmatrix}^T \in R^m$.

By assuming that $\mathbf{\Lambda}(k) = \mathbf{0}$, an $m^{th}$-order observer can be synthesised for the auxiliary variable vector $\mathbf{z}(k)$ as follows:

$$\hat{\mathbf{z}}(k+1) = \mathbf{\Gamma \hat{z}}(k) + \mathbf{\Omega_x x}(k) + \mathbf{\Omega_u} u(k) \quad (6)$$

where $\hat{\mathbf{z}}(k) = \begin{bmatrix} \hat{z}_0(k) & \cdots & \hat{z}_m(k) \end{bmatrix}^T$ is the estimated $\mathbf{z}(k)$ at $kT_s$ seconds.

Section III shows that a uniformly ultimately bounded and a uniformly asymptotically stable observers can be synthesised for the auxiliary variable $\mathbf{z}(k)$ when $\mathbf{\Lambda}(k)$ is bounded and null, respectively. Hence, the disturbance variable and its derivatives up to order $m$ can be obtained by substituting Eq. (6) into Eq. (4). Let us now synthesise the conventional ZO and First-Order (FO) DObs using the proposed DOb synthesis method.

*B1. Conventional Zero-Order DOb:*

When the truncation error $R_0(k) = \dot{\tau}_{dn}(\xi_k)(T_s - \lambda)$ is neglected in the zero-order Taylor series expansion of the discrete disturbance vector, the approximate dynamic model of the servo system can be derived as follows:

$$\mathbf{x}(k+1) = \mathbf{A_{Dn}} \mathbf{x}(k) + \mathbf{B_{Dn}} u(k) - \mathbf{D_{Dn}} \tau_{dn}(k) \quad (7)$$

where $\mathbf{D_{Dn}} = \int_0^{T_s} e^{\mathbf{A_{Cn}}\lambda} d\lambda \mathbf{D_{Cn}}$ is the discrete disturbance input vector.

To estimate $\tau_{dn}(k)$, an auxiliary variable, as expressed in Eq. (8), is introduced.

$$z_0(k) = \tau_{dn}(k) + \mathbf{L_0}^T \mathbf{x}(k) \quad (8)$$

The dynamics of the auxiliary variable $z_0(k)$ is derived by substituting Eq. (7) into Eq. (8) as follows:

$$\mathbf{z}(k+1) = \mathbf{\Gamma z}(k) + \mathbf{\Omega_x x}(k) + \mathbf{\Omega_u} u(k) + \mathbf{\Lambda}(k) \quad (9)$$

where $\mathbf{z}(k) = z_0(k) \in R$, $\mathbf{\Gamma} = (1 - \mathbf{L_0}^T \mathbf{D_{Dn}}) \in R$ is the state transition coefficient, $\mathbf{\Omega_x} = \mathbf{L_0}^T (\mathbf{A_{Dn}} + \mathbf{D_{Dn}} \mathbf{L_0}^T - \mathbf{I}) \in R^{1 \times 2}$ is the coefficient matrix of $\mathbf{x}(k)$, $\mathbf{\Omega_u} = \mathbf{L_0}^T \mathbf{B_{Dn}} \in R$ is the control input coefficient, and $\mathbf{\Lambda}(k) = \tau_{dn}(k+1) - \tau_{dn}(k) \in R$ is the variation of the nominal disturbance variable between $kT_s$ and $(k+1)T_s$ seconds. The ZO DOb can be synthesised by substituting Eq. (9) into Eq. (6) and neglecting the term $\mathbf{\Lambda}(k)$. Hence, the disturbance variable is estimated using $\hat{\tau}_{dn}(k) = \hat{z}_0(k) - \mathbf{L_0}^T \mathbf{x}(k)$ where $\hat{\tau}_{dn}(k)$ represents the estimated $\tau_{dn}(k)$ at $kT_s$ seconds.

*B2. First-Order DOb:*

The approximate dynamic model of the servo system, derived by neglecting the truncation error of the disturbance variable's first-order Taylor series expansion $R_1(k)$, is given by:

$$\mathbf{x}(k+1) = \mathbf{A_{Dn}} \mathbf{x}(k) + \mathbf{B_{Dn}} u(k) - \mathbf{D_{Dn}} \tau_{dn}(k) - \tilde{\mathbf{D}}_{\mathbf{Dn}} \dot{\tau}_{dn}(k) \quad (10)$$

where $\mathbf{D_{Dn}} = \int_0^{T_s} e^{\mathbf{A_{Cn}}\lambda} d\lambda \mathbf{D_{Cn}}$ and $\tilde{\mathbf{D}}_{\mathbf{Dn}} = \int_0^{T_s} e^{\mathbf{A_{Cn}}\lambda} (T_s - \lambda) d\lambda \mathbf{D_{Cn}}$ are the disturbance input vectors, and $R_1(k) = (1/2) \ddot{\tau}_{dn}(\xi_k)(T_s - \lambda)^2$ is the truncation error of order 2.

Equation (11) represents the auxiliary variables utilised in the FO DOb synthesis.

$$\begin{aligned} z_0(k) &= \tau_{dn}(k) + \mathbf{L_0}^T \mathbf{x}(k) \\ z_1(k) &= \dot{\tau}_{dn}(k) + \mathbf{L_1}^T \mathbf{x}(k) \end{aligned} \quad (11)$$

The dynamics of the auxiliary variables can be derived by substituting Eq. (10) into Eq. (11) as follows:

$$\mathbf{z}(k+1) = \mathbf{\Gamma z}(k) + \mathbf{\Omega_x x}(k) + \mathbf{\Omega_u} u(k) + \mathbf{\Lambda}(k) \quad (12)$$

where $\mathbf{z}(k) = \begin{bmatrix} z_0(k) & z_1(k) \end{bmatrix}^T$ is the auxiliary variable vector,

$\mathbf{\Gamma} = \begin{bmatrix} 1 - \mathbf{L_0}^T \mathbf{D_{Dn}} & T_s - \mathbf{L_0}^T \tilde{\mathbf{D}}_{\mathbf{Dn}} \\ -\mathbf{L_1}^T \mathbf{D_{Dn}} & 1 - \mathbf{L_1}^T \tilde{\mathbf{D}}_{\mathbf{Dn}} \end{bmatrix} \in R^{2 \times 2}$ is the auxiliary variable state matrix, $\mathbf{\Omega_x} = \begin{bmatrix} \mathbf{L_0}^T \mathbf{A_{Dn}} - (1 - \mathbf{L_0}^T \mathbf{D_{Dn}}) \mathbf{L_0}^T - (T_s - \mathbf{L_0}^T \tilde{\mathbf{D}}_{\mathbf{Dn}}) \mathbf{L_1}^T \\ \mathbf{L_1}^T \mathbf{A_{Dn}} + \mathbf{L_1}^T \mathbf{D_{Dn}} \mathbf{L_0}^T - (1 - \mathbf{L_1}^T \tilde{\mathbf{D}}_{\mathbf{Dn}}) \mathbf{L_1}^T \end{bmatrix} \in R^{2 \times 2}$ and $\mathbf{\Omega_u} = \begin{bmatrix} \mathbf{L_0}^T \mathbf{B_{Dn}} \\ \mathbf{L_1}^T \mathbf{B_{Dn}} \end{bmatrix} \in R^2$ are the coefficient matrices of $\mathbf{x}(k)$ and control input vector, respectively, and $\mathbf{\Lambda}(k) = \begin{bmatrix} 0 & \dot{\tau}_{dn}(k+1) - \dot{\tau}_{dn}(k) \end{bmatrix}^T \in R^2$.

The FO DOb can be similarly synthesised by substituting Eq. (12) into Eq. (6) and assuming that $\mathbf{\Lambda}(k)$ is null. The estimated disturbance and its first-order derivative can be obtained using $\hat{\tau}_{dn}(k) = \hat{z}_0(k) - \mathbf{L_0}^T \mathbf{x}(k)$ and $\hat{\dot{\tau}}_{dn}(k) = \hat{z}_1(k) - \mathbf{L_1}^T \mathbf{x}(k)$, respectively.

### C. A High Performance DOb Synthesis:

Since a static disturbance model is utilised in the synthesis of the ZO DOb, its disturbance estimation accuracy is inherently limited by the first-order truncation error described in Eq. (13).

$$\Delta^+(k) = \tau_{dn}(k+1) - \tau_{dn}(k) = \dot{\tau}_{dn}(k)T_s + 1/2\ddot{\tau}_{dn}(\xi_k)T_s^2 \quad (13)$$

This limitation may result in compromised motion control performance when servo systems are exposed to dynamic disturbances, a condition commonly encountered in real-world applications [1–3, 11–14]. While the estimation accuracy can be improved by reducing the truncation error in a HO DOb (e.g., the second order truncation error of the FO DOb in Eq. (10)), the robust motion controller becomes increasingly sensitive to noise due to the incorporation of higher-order derivatives of the disturbance variable, as shown in Eqs. (4) and (11) [19].

To reduce disturbance estimation error while relying solely on estimated disturbances (without incorporating the derivatives of disturbances) in the robust motion controller, let us consider the following truncation error, which is derived from the backward Taylor series expansion of the nominal disturbance variable.

$$\Delta^-(k) = \tau_{dn}(k-1) - \tau_{dn}(k) = -\dot{\tau}_{dn}(k)T_s + 1/2\ddot{\tau}_{dn}(\xi_{k-1})T_s^2 \quad (14)$$

By summing Eqs. (13) and (14), a truncation error of second order is obtained, as expressed in Eq. (15).

$$\Delta\tau_{dn}(k) = \Delta^+\tau_{dn}(k) + \Delta^-\tau_{dn}(k) = \\ \tau_{dn}(k+1) - 2\tau_{dn}(k) + \tau_{dn}(k-1) = 1/2\big(\ddot{\tau}_{dn}(\xi_k) + \ddot{\tau}_{dn}(\xi_{k-1})\big)T_s^2 \quad (15)$$

To synthesize the HP DOb with the truncation error described in Eq. (15), let us introduce the following auxiliary variables.

$$\begin{aligned} z_0(k) &= \tau_{dn}(k-1) + \mathbf{L}_0^T \mathbf{x}(k) \\ z_1(k) &= \tau_{dn}(k) + \mathbf{L}_1^T \mathbf{x}(k) \end{aligned} \quad (16)$$

where $\mathbf{L}_0$ and $\mathbf{L}_1 \in R^2$ are the observer gain of the HP DOb.

The dynamics of the auxiliary variables can be derived by substituting Eq. (7) into Eq. (16) as follows:

$$\mathbf{z}(k+1) = \mathbf{\Gamma z}(k) + \mathbf{\Omega}_\mathbf{x}\mathbf{x}(k) + \mathbf{\Omega}_\mathbf{u}u(k) + \mathbf{\Lambda}(k) \quad (17)$$

where $\mathbf{z}(k) = [z_0(k) \ z_1(k)]^T$, $\mathbf{\Gamma} = \begin{bmatrix} 0 & 1 - \mathbf{L}_0^T\mathbf{D}_{\mathbf{Dn}} \\ -1 & 2 - \mathbf{L}_1^T\mathbf{D}_{\mathbf{Dn}} \end{bmatrix}$, $\mathbf{\Omega}_\mathbf{u} = \begin{bmatrix} \mathbf{L}_0^T\mathbf{B}_{\mathbf{Dn}} \\ \mathbf{L}_1^T\mathbf{B}_{\mathbf{Dn}} \end{bmatrix}$,

$\mathbf{\Omega}_\mathbf{x} = \begin{bmatrix} \mathbf{L}_0^T\mathbf{A}_{\mathbf{Dn}} - \big(1 - \mathbf{L}_0^T\mathbf{D}_{\mathbf{Dn}}\big)\mathbf{L}_1^T \\ \mathbf{L}_0^T + \mathbf{L}_1^T\mathbf{A}_{\mathbf{Dn}} - \big(2 - \mathbf{L}_1^T\mathbf{D}_{\mathbf{Dn}}\big)\mathbf{L}_1^T \end{bmatrix}$, and $\mathbf{\Lambda}(k) = \begin{bmatrix} 0 & \Delta\tau_{dn}(k) \end{bmatrix}^T$.

By neglecting the truncation error $\Delta\tau_{dn}(k)$, the dynamic model of the HP DOb is obtained as expressed in Eq. (18).

$$\hat{\mathbf{z}}(k+1) = \mathbf{\Gamma}\hat{\mathbf{z}}(k) + \mathbf{\Omega}_\mathbf{x}\mathbf{x}(k) + \mathbf{\Omega}_\mathbf{u}u(k) \quad (18)$$

where $\hat{\mathbf{z}}(k) = [\hat{z}_0(k) \ \hat{z}_1(k)]^T$ represents the estimated auxiliary variable vector at $kT_s$ seconds. The nominal disturbance variable is estimated using $\hat{\tau}_{dn}(k) = \hat{z}_1(k) - \mathbf{L}_1^T\mathbf{x}(k)$.

### III. DOb Analysis in the Discrete-time Domain

The estimation error dynamics of the proposed DObs can be derived by subtracting $\hat{\mathbf{z}}(k+1)$ from $\mathbf{z}(k+1)$ as follows:

$$\mathbf{e}_\mathbf{z}(k+1) = \mathbf{\Gamma}\mathbf{e}_\mathbf{z}(k) + \mathbf{\Lambda}(k) \quad (19)$$

where $\mathbf{e}_\mathbf{z}(k) = \mathbf{z}(k) - \hat{\mathbf{z}}(k)$ is the estimation error vector, $\mathbf{\Gamma}$ and $\mathbf{\Lambda}(k)$ are the state transition matrix and truncation error vector given in Eqs. (9), (12) and (17) for the conventional ZO, FO, and HP DObs, respectively.

To analyse the stability of the proposed DObs, let us consider the Lyapunov function candidate given by:

$$V(k) = \mathbf{e}_\mathbf{z}^T(k)\mathbf{P}\mathbf{e}_\mathbf{z}(k) \quad (20)$$

where $\mathbf{P}$ is a symmetric positive definite matrix.

The forward difference of the Lyapunov function candidate can be derived by substituting Eq. (19) into Eq. (20) as follows:

$$\Delta V(k) = -\mathbf{e}_\mathbf{z}^T(k)\mathbf{Q}\mathbf{e}_\mathbf{z}(k) + 2\mathbf{e}_\mathbf{z}^T(k)\mathbf{\Gamma}^T\mathbf{P}\mathbf{\Lambda}(k) + \mathbf{\Lambda}^T(k)\mathbf{P}\mathbf{\Lambda}(k) \quad (21)$$

where $\Delta V(k) = V(k+1) - V(k)$, and $\mathbf{Q} = \mathbf{P} - \mathbf{\Gamma}^T\mathbf{P}\mathbf{\Gamma}$ is a symmetric positive definite matrix.

When it is assumed that the disturbance vector is bounded by $\|\mathbf{\Lambda}(k)\| \leq d_k$ where $d_k > 0 \in R$, the following inequality holds.

$$\begin{aligned} \Delta V(k) &\leq -\mathbf{e}_\mathbf{z}^T(k)\mathbf{Q}\mathbf{e}_\mathbf{z}(k) + \|\mathbf{e}_\mathbf{z}(k)\|^2 + \|\mathbf{\Gamma}^T\mathbf{P}\mathbf{\Lambda}(k)\|^2 + \mathbf{\Lambda}^T(k)\mathbf{P}\mathbf{\Lambda}(k) \\ &\leq -\kappa_e\|\mathbf{e}_\mathbf{z}(k)\|^2 + \kappa_d\|d_k\|^2 \end{aligned} \quad (22)$$

where $\kappa_e = \lambda_{\min}(\mathbf{Q}) - 1$ and $\kappa_d = \lambda_{\max}^2(\mathbf{\Gamma}^T\mathbf{P}) + \lambda_{\max}(\mathbf{P})$ in which $\lambda_{\min}(\bullet)$ and $\lambda_{\max}(\bullet)$ represent the minimum and maximum eigenvalues of matrix $\bullet$, respectively.

Equation (22) shows that $\Delta V(k)$ is smaller than zero when $\kappa_e > 0$ and $\|\mathbf{e}_\mathbf{z}(k)\| > \sqrt{\kappa_d/\kappa_e}\|d_k\|$. Therefore, any estimation error originating outside the set $\Phi_S = \{\mathbf{e}_\mathbf{z}(k) \in R^m : \|\mathbf{e}_\mathbf{z}(k)\| \leq \sqrt{\kappa_d/\kappa_e}\|d_k\|\}$ will ultimately converge within the set $\Phi_S$, ensuring the uniformly ultimate boundedness. When $d_k$ is null, $\Delta V(k) < 0$ for all $\kappa_e > 0$, ensuring the asymptotic stability of the digital DObs.

Let us now tune the observer gain vectors of the ZO, FO, and HP DObs by using the proposed Lyapunov stability analysis.

### A. Conventional Zero-Order DOb:

Let us employ the observer gain vector as specified in Eq. (23) for the ZO DOb.

$$\mathbf{L}_\mathbf{0} = \big(\ell_0/|\mathbf{D}_{\mathbf{Dn}}|_1\big)\mathbf{v} \quad (23)$$

where $\ell_0 \in R$ is a free control parameter, $|\bullet|_1$ is the L1-norm of $\bullet$ and $\mathbf{v} = [1 \ 1]^T$. Substituting Eqs. (9) and (23) into Eq. (21) yields

$$\mathbf{Q} = \mathbf{P}\ell_0(2 - \ell_0) \quad (24)$$

where $\mathbf{P} \in R$ is a positive scalar. To achieve Lyapunov stability, $\mathbf{Q} \in R$ should be higher than zero, hence $0 < \ell_0 < 2$.

### B. First-Order DOb:

Let us employ the following observer gain vectors in the FO DOb synthesis.

$$\mathbf{L}_\mathbf{0} = \ell_0\mathbf{v} \text{ and } \mathbf{L}_\mathbf{1} = \ell_1\mathbf{v} \quad (25)$$

where $\ell_0$ and $\ell_1 \in R$ are the free control parameters of the FO DOb, and $\mathbf{v} = [6/D_{Dn1} \ -4/D_{Dn2}]^T$ in which $\mathbf{D}_{\mathbf{Dn}} = [D_{Dn1} \ D_{Dn2}]^T$.

Substituting Eqs. (12) and (25) into Eq. (19) yields

$$\mathbf{e_z}(k+1) = \mathbf{\Gamma e_z}(k) + \mathbf{\Lambda}(k) \quad (26)$$

where $\mathbf{e_z}(k) = \begin{bmatrix} z_0 - \hat{z}_0 & z_1 - \hat{z}_1 \end{bmatrix}^T$ and $\mathbf{\Gamma} = \begin{bmatrix} 1-2\ell_0 & T_s \\ -2\ell_1 & 1 \end{bmatrix}$.

The error dynamics in Eq. (26) can be rewritten using the Jordan canonical form of $\mathbf{\Gamma}$ as follows:

$$\tilde{\mathbf{e}}_{\mathbf{z}}(k+1) = \tilde{\mathbf{\Gamma}} \tilde{\mathbf{e}}_{\mathbf{z}}(k) + \tilde{\mathbf{\Lambda}}(k) \quad (27)$$

where $\tilde{\mathbf{\Gamma}} = \mathbf{S}^{-1} \mathbf{\Gamma} \mathbf{S} = \begin{bmatrix} \lambda_1 & 0 \\ 0 & \lambda_2 \end{bmatrix}$ is a similar matrix to $\mathbf{\Gamma}$ in which $\lambda_{1,2} = 1 - \ell_0 \mp \sqrt{\ell_0^2 - 2\ell_1 T_s}$, $\mathbf{S} = \begin{bmatrix} 1-\lambda_1/2\ell_1 & 1-\lambda_2/2\ell_1 \\ 1 & 1 \end{bmatrix}$ is a change-of-basis matrix, $\tilde{\mathbf{e}}_{\mathbf{z}}(k) = \mathbf{S}^{-1} \mathbf{e}_{\mathbf{z}}(k)$ and $\tilde{\mathbf{\Lambda}}(k) = \mathbf{S}^{-1} \mathbf{\Lambda}(k)$.

Substituting Eq. (27) into Eq. (20) yields

$$\mathbf{Q} = \mathbf{P} - \tilde{\mathbf{\Gamma}}^T \mathbf{P} \tilde{\mathbf{\Gamma}} = \begin{bmatrix} (1-\lambda_1^2) p_{11} & (1-\lambda_1 \lambda_2) p_{12} \\ (1-\lambda_1 \lambda_2) p_{12} & (1-\lambda_2^2) p_{22} \end{bmatrix} \quad (28)$$

where $p_{ij}$ represents the $i^{th}$ row and $j^{th}$ column of the symmetric positive definite $\mathbf{P}$ matrix in the Lyapunov function candidate. The necessary and sufficient conditions for a positive definite $\mathbf{Q}$ matrix are as follows:

$$(1-\lambda_1^2) p_{11} > 0, \ (1-\lambda_2^2) p_{22} > 0,$$
$$p_{11} p_{22} (1-\lambda_1^2)(1-\lambda_2^2) - p_{12}^2 (1-\lambda_1 \lambda_2) > 0, \quad (29)$$

Since $\mathbf{P} > 0$, $p_{11} > 0$, $p_{22} > 0$, and $p_{11} p_{22} > p_{12}^2$. Hence, the stability constraints on the free control parameters of the FO DOb are derived from Eq. (29) as follows:

$$|\lambda_{1,2}| = \left|1 - \ell_0 \mp \sqrt{\ell_0^2 - 2\ell_1 T_s}\right| < 1 \quad (30)$$

From Eq. (30), the observer gain vectors can be tuned using $\ell_0 = (1/2)(2 + \lambda_1^{des} + \lambda_2^{des})$ and $\ell_1 = (1/2T_s)(\lambda_1^{des} \lambda_2^{des} + 2\ell_0 - 1)$ where $\lambda_1^{des}$ and $\lambda_2^{des}$ represent the desired eigenvalues of the FO DOb's error dynamics.

### C. High-Performance DOb:

Last, let us synthesise the HP DOb by using the observer gain vectors given by

$$\mathbf{L_0} = (0.5 + \ell_0) \mathbf{v} \text{ and } \mathbf{L_1} = (1 + \ell_1) \mathbf{v} \quad (31)$$

where $\ell_0$ and $\ell_1 \in R$ are the free control parameters of the HP DOb, and $\mathbf{v} = \begin{bmatrix} 1/D_{Dn1} & 1/D_{Dn2} \end{bmatrix}^T$ in which $\mathbf{D_{Dn}} = \begin{bmatrix} D_{Dn1} & D_{Dn2} \end{bmatrix}^T$.

By using the Jordan canonical form, the error dynamics of the HP DOb can be derived as follows:

$$\tilde{\mathbf{e}}_{\mathbf{z}}(k+1) = \tilde{\mathbf{\Gamma}} \tilde{\mathbf{e}}_{\mathbf{z}}(k) + \tilde{\mathbf{\Lambda}}(k) \quad (32)$$

where $\tilde{\mathbf{\Gamma}} = \mathbf{S}^{-1} \mathbf{\Gamma} \mathbf{S} = \begin{bmatrix} \lambda_1 & 0 \\ 0 & \lambda_2 \end{bmatrix}$ is a similar matrix to $\mathbf{\Gamma} = \begin{bmatrix} 0 & -2\ell_0 \\ -1 & -2\ell_1 \end{bmatrix}$ in Eq. (17), $\mathbf{S} = \begin{bmatrix} \lambda_2 & \lambda_1 \\ 1 & 1 \end{bmatrix}$ is a change-of-basis matrix in which $\lambda_{1,2} = -\ell_1 \mp \sqrt{2\ell_0 + \ell_1^2}$, $\tilde{\mathbf{e}}_{\mathbf{z}}(k) = \mathbf{S}^{-1} \mathbf{e}_{\mathbf{z}}(k)$ and $\tilde{\mathbf{\Lambda}}(k) = \mathbf{S}^{-1} \mathbf{\Lambda}(k)$.

Substituting Eq. (32) into Eq. (20) yields the same positive definite $\mathbf{Q}$ matrix and necessary and sufficient conditions given in Eqs. (28) and (29). Hence, the stability constraints on the free control parameters $\ell_0$ and $\ell_1$ can be similarly derived as follows:

$$|\lambda_{1,2}| = \left|-\ell_1 \mp \sqrt{2\ell_0 + \ell_1^2}\right| < 1 \quad (33)$$

From Eq. (33), the observer gain vectors can be similarly tuned using $\ell_0 = (-1/2)(\lambda_1^{des} \lambda_2^{des})$ and $\ell_1 = (1/2)(\lambda_1^{des} + \lambda_2^{des})$ where $\lambda_1^{des}$ and $\lambda_2^{des}$ represent the desired eigenvalues of the HP DOb's error dynamics.

## IV. DOb-Based Digital Robust Motion Controller

The DOb-based digital robust motion controller is illustrated in Fig. 1. It is implemented using the following inner- and outer-feedback control loops.

### A. Inner-Feedback Control Loop Analysis and Synthesis:

To achieve robustness, the estimated nominal disturbance variable is fed back in an inner-loop, as illustrated in Fig. 1. However, the inner-loop feedback control changes the free control parameters' stability constraints derived in Eqs. (24), (30) and (33). Let us first analyse the stability of the digital robust motion control system in the inner-loop.

When $\hat{\tau}_{dn}(k)$ is fed back using the conventional ZO DOb in the inner-loop, an augmented state-space model can be obtained for the digital robust motion controller as follows:

$$\boldsymbol{\xi}(k+1) = \mathbf{A_{DI}} \boldsymbol{\xi}(k) + \mathbf{B_{DI}} u_p(k) - \mathbf{\Pi_{DI}} \quad (34)$$

where $\boldsymbol{\xi}(k) = \begin{bmatrix} \mathbf{x}(k) & z_0(k) \end{bmatrix}^T$, $\mathbf{A_{DI}} = \begin{bmatrix} \mathbf{A_D} - \mathbf{B_D} \mathbf{L_0}^T & \mathbf{B_D} \\ \mathbf{L_0}^T (\mathbf{A_{Dn}} - \mathbf{I_2}) & 1 \end{bmatrix}$ in which $\mathbf{I_2}$ is a 2x2 identity matrix, $\mathbf{B_{DI}} = \begin{bmatrix} \mathbf{B_D}^T & \mathbf{L_0}^T \mathbf{B_{Dn}} \end{bmatrix}^T$, $\mathbf{\Pi_{DI}} = \begin{bmatrix} \mathbf{\Pi_D}^T & 0 \end{bmatrix}^T$ and $u(k) = u_p(k) + \hat{\tau}_{dn}(k) = u_p(k) + z_0(k) - \mathbf{L_0}^T \mathbf{x}(k)$ in which $u_p(k)$ is the performance control input at $kT_s$ seconds.

Equation (34) can be rewritten using the Jordan canonical form of the augmented state transition matrix as follows:

$$\tilde{\boldsymbol{\xi}}(k+1) = \tilde{\mathbf{A}}_{\mathbf{DI}} \tilde{\boldsymbol{\xi}}(k) + \tilde{\mathbf{B}}_{\mathbf{DI}} u_p(k) - \tilde{\mathbf{\Pi}}_{\mathbf{DI}} \quad (35)$$

where $\tilde{\mathbf{A}}_{\mathbf{DI}} = \mathbf{S}^{-1} \mathbf{A_{DI}} \mathbf{S} = \begin{bmatrix} 1 & 0 & 0 \\ 0 & e^{-bT_s} & 0 \\ 0 & 0 & 1-\alpha \ell_0 \end{bmatrix}$ is a similar matrix to $\mathbf{A_{DI}}$ in which $\alpha = J_{mn}/J_m$ and $b = b_m/J_m = b_{mn}/J_{mn}$ is a viscous friction term in the servo system model; $\mathbf{S} = \begin{bmatrix} \rho_1(1-1/b) & -\rho_1/b & \rho_1\rho_2\rho_3 \\ 0 & \rho_1 & \rho_1\rho_2 \\ 1 & 1 & 1 \end{bmatrix}$ is a change-of-basis matrix in which $\rho_1 = \frac{1-b-bT_s+e^{-bT_s}(b-1)}{b(b-1)J_{mn}\ell_0}$, $\rho_2 = \frac{J_{mn}\ell_0}{(1-e^{-bT_s})J_m}$, and $\rho_3 = \left(1 - \frac{1}{bT_s} + \frac{e^{-bT_s}}{1-e^{-bT_s}} - \frac{J_m}{J_{mn}\ell_0}\right)T_s$; $\tilde{\boldsymbol{\xi}}(k) = \mathbf{S}^{-1}\boldsymbol{\xi}(k)$, $\tilde{\mathbf{B}}_{DI} = \mathbf{S}^{-1}\mathbf{B}_{DI}$, and $\tilde{\boldsymbol{\Pi}}_{DI} = \mathbf{S}^{-1}\boldsymbol{\Pi}_{DI}$.

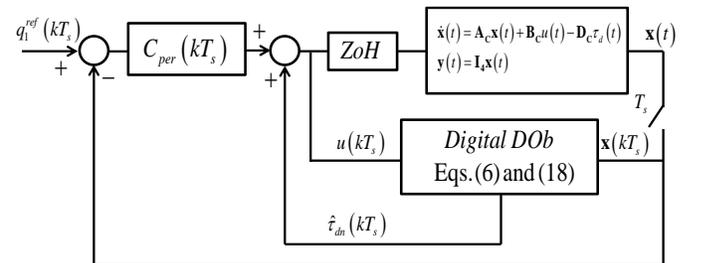

Fig. 1: DOb-based digital robust motion controller. $C_{per}(kT_s)$ represents the outer-loop performance controller, which is implemented using a PD controller in this paper. $q_1^{ref}(kT_s)$ represents the position reference of Motor 1.

The Lyapunov stability analysis of the closed-loop system is derived using Eq. (35) as follows:

$$V(k) = \tilde{\boldsymbol{\xi}}(k)\mathbf{P}\tilde{\boldsymbol{\xi}}(k) \tag{36}$$

$$\Delta V(k) \leq -\kappa_\xi \|\tilde{\boldsymbol{\xi}}(k)\|^2 + \kappa_\Pi \|\tilde{\boldsymbol{\Pi}}_{DI}(k)\|^2 + \kappa_u \|u_p(k)\|^2 \tag{37}$$

where $\kappa_\xi = \lambda_{\min}(\mathbf{Q}) - 1$, $\kappa_u = \left(|\lambda_{\max}(\mathbf{P})| + \lambda_{\max}^2(\mathbf{P}) + \lambda_{\max}^2(\tilde{\mathbf{A}}_{DI}^T\mathbf{P})\right)\|\tilde{\mathbf{B}}_{DI}\|^2$, $\kappa_\Pi = 1 + |\lambda_{\max}(\mathbf{P})| + \lambda_{\max}^2(\tilde{\mathbf{A}}_{DI}^T\mathbf{P})$, $\mathbf{P} \in R^{3\times 3}$ is a diagonal positive definite matrix, and

$$\mathbf{Q} = \mathbf{P} - \tilde{\mathbf{A}}_{DI}^T \mathbf{P} \tilde{\mathbf{A}}_{DI} = \begin{bmatrix} 0 & 0 & 0 \\ 0 & p_{22}(1-e^{-2bT_s}) & 0 \\ 0 & 0 & p_{33}(1-(1-\alpha\ell_0)^2) \end{bmatrix} \tag{38}$$

where $p_{ii} > 0$ represents the diagonal term of the $\mathbf{P}$ matrix.

To obtain a positive semi-definite $\mathbf{Q}$ matrix, the stability constraint on the free control parameter $\ell_0$ is follows:

$$0 < \alpha\ell_0 < 2 \tag{39}$$

Equations (24) and (39) show that the stability constraint on the free control parameter of the DOb changes when the estimated disturbance is fed back in the inner-loop. The positive semi-definite $\mathbf{Q}$ matrix refers to the marginal stability of the inner-loop. This, however, does not lead to a practical problem, because the velocity state of the servo system goes to zero while the position state remains fixed at a new equilibrium point after a disturbance is applied to the servo system when $0 < \alpha\ell_0 < 2$.

### B. Outer-Feedback Control Loop Analysis and Synthesis:

To adjust the performance of the servo system, let us employ the following state feedback controller in the outer-loop.

$$u_p(k) = q_1^{ref}(k) - \tilde{\mathbf{K}}^T\tilde{\boldsymbol{\xi}}(k) \tag{40}$$

where $\tilde{\mathbf{K}} = [K_1 \quad K_2 \quad 0]^T \in R^3$ is the control gain vector in which $K_1$ and $K_2 \in R$ are free control parameters.

Substituting Eq. (40) into Eq. (35) yields

$$\tilde{\boldsymbol{\xi}}(k+1) = \tilde{\mathbf{A}}_{DO}\tilde{\boldsymbol{\xi}}(k) + \tilde{\mathbf{B}}_{DI}q_1^{ref}(k) - \tilde{\boldsymbol{\Pi}}_{DI} \tag{41}$$

where $\tilde{\mathbf{A}}_{DO} = \tilde{\mathbf{A}}_{DI} - \tilde{\mathbf{B}}_{DI}\tilde{\mathbf{K}}^T = \begin{bmatrix} \mathbf{a}_{11} & \mathbf{a}_{12} \\ \mathbf{a}_{21} & 1-\alpha\ell_0 \end{bmatrix} \in R^{3\times 3}$ is the closed-loop state matrix in which $\mathbf{a}_{11} = \begin{bmatrix} 1-K_1(\ell_0+\delta_1) & 1-K_2(\ell_0+\delta_1) \\ -2K_1\delta_2 & 1-2K_2\delta_2 \end{bmatrix} \in R^{2\times 2}$, $\mathbf{a}_{12} = \mathbf{0} \in R^2$, $\mathbf{a}_{21} = \begin{bmatrix} K_1\delta_1 \\ K_2\delta_1 \end{bmatrix}^T \in R^{1\times 2}$, $\delta_1 = \frac{2(J_{mn}-J_m)T_s}{J_{mn}(T_s+2)}$, and $\delta_2 = \frac{2\ell_0 T_s}{(T_s+2)}$.

Since the off-diagonal term $\mathbf{a}_{12}$ is null, the characteristic function of $\tilde{\mathbf{A}}_{DO}$ is as follows:

$$|\lambda \mathbf{I}_3 - \tilde{\mathbf{A}}_{DO}| = |\lambda - (1-\alpha\ell_0)||\lambda \mathbf{I}_2 - \mathbf{a}_{11}| \tag{42}$$

where $\mathbf{I}_*$ is a $*\times*$ identity matrix.

As shown in Eq. (42), the gains of the outer-loop controller can be independently tuned to adjust the performance of the digital motion control system while the inner-loop's eigenvalue $1-\alpha\ell_0$ is kept fixed. Nevertheless, $\mathbf{a}_{11}$ and $\mathbf{a}_{21}$ matrices show that the observer gain $\ell_0$ and the performance control gains $K_1$ and $K_2$ still affect the robustness and performance of the DOb-based digital robust motion controller. Once the eigenvalues of $\mathbf{a}_{11}$ is tuned in Eq. (42), the state feedback controller can be obtained using $\mathbf{K} = \tilde{\mathbf{K}}\mathbf{S}^{-1}$.

Similarly, the constraints on the key design parameters of the HO and HP DObs must be derived when estimated disturbances are fed back in the inner-loop. As the order of DOb increases, employing numerical methods may become necessary to efficiently perform these derivations and analyses. For brevity, these detailed derivations have been omitted from this paper.

## V. EXPERIMENTS

### a) Experimental Procedure

The servo system illustrated in Fig. 2 was used to validate the proposed analysis and synthesis methods. It was built using two Electronically Communicated (EC) flat motors (Maxon EC 90: 607931), ESCON 70/10 motor drivers, Omron's E6B2 rotary encoders with 2500 pulse per revolution, and Meßsysteme's K6D27 torque sensor. The real-time motion control algorithms were implemented using Matlab 2022b and Quanser QPIDe data acquisition card along with QUARC 2023 (4.3.4128) software. To apply external disturbances, the shafts of the motors were attached each other using a coupler as shown in the figure. The position of the first motor (Motor 1) was controlled using different motion controllers while external disturbances were exerted via the second motor (Motor 2). The disturbance torque was estimated using the K6D27 torque sensor and reaction force observer [22]. The sampling time was set to 1ms in the experiments.

To evaluate the stability and performance of the proposed robust motion controllers, regulation and trajectory tracking control experiments were performed using step and sinusoidal reference inputs with an amplitude of $\pi/2$ rad. Concurrently, Motor 2 was controlled using a DOb-based robust torque controller with a proportional torque control gain $C_\tau = 0.1$.

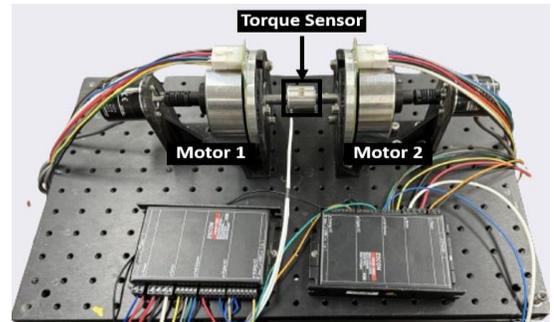

Fig. 2: Experimental setup.

This setup allowed Motor 2 to apply the disturbance torque specified in Eq. (45) to the shaft of Motor 1 over the time interval from $t = 3$ seconds to $t = 8$ seconds.

$$\tau_2^{ref}(k) = 0.35\sin(2.5\pi(t-3)) + 0.47\cos(1.7\pi(t-3)) + 0.56\sin(1.5\pi(t-3))\cos(3.5\pi(t-3)) \quad (43)$$

*b) Regulation Control Performance of Motion Controllers:*

Figure 3 illustrates the regulation control results obtained when the following motion controllers were employed in the experiments:

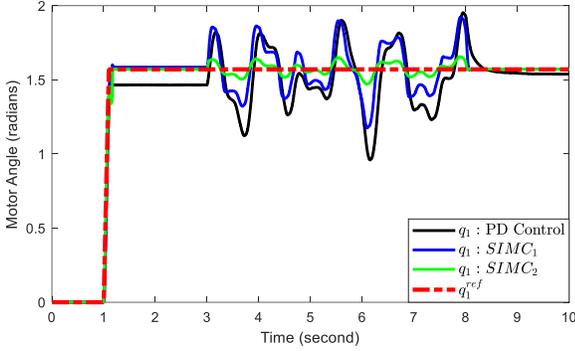

a) PD and SIMC-based motion controllers. In PD controller, $K_p = 2.5$ and $K_d = 0.25$. The free control parameter of SIMC was $\tau_c = 0.005$ and $\tau_c = 0.0025$ for $SIMC_1$ and $SIMC_2$, respectively.

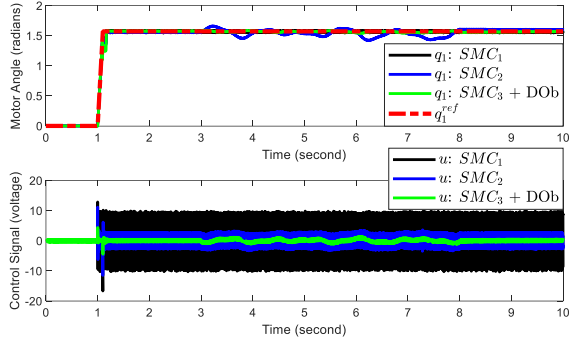

b) SMC based robust motion controller. The SMC gains for $SMC_1$, $SMC_2$ and $SMC_3$ were 10, 1 and 0.1, respectively, while the sliding surface gain was set to 1. The bandwidth of the DOb was set to 350 rad/s.

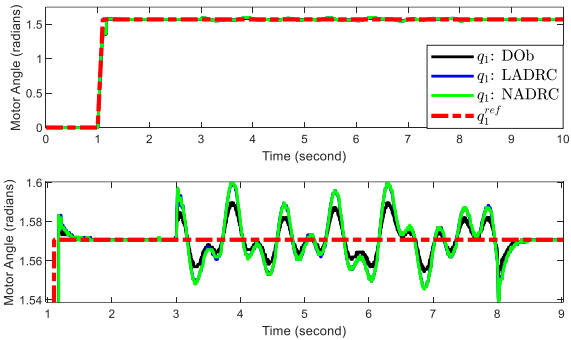

c) Observer-based robust motion controllers. The bandwidths of the DOb and LADRC were set to 350 rad/s, and the NADRC were tuned using the nonlinear function $fal(\hat{\alpha}, \delta)$ proposed in [13], where $\hat{\alpha} = 0.001$ and $\delta = 0.95$. The outer-loop PD controller gains were $K_p = 2.5$ and $K_d = 0.25$.

Fig. 3: Regulation control of Motor 1 using i) PD controller, ii) SIMC, iii) SMC and iv) DOb, LADRC and NADRC-based controllers. $q_1$ and $q_1^{ref}$ represent the position state of Motor 1 and its reference in regulation control, respectively.

1) Conventional Proportional Derivative (PD) control with the proportional and derivative control gains $K_p = 2.5$ and $K_d = 0.25$, respectively.
2) Simple Internal Model Control (SIMC) when the robust motion controller was tuned using $\tau_c = 0.005$ and $\tau_c = 0.0025$, where $\tau_c$ is the desired closed-loop time constant employed for controller tuning, as described in [31].
3) Sliding Mode Control (SMC): The SMC gains were 0.1, 1 and 10 while the sliding surface gain was 1 [32, 33].
4) DOb-based Robust Motion Control: i) Conventional DOb [12], ii) Linear ADRC (LADRC) [13], and iii) Nonlinear ADRC (NADRC) tuned using the nonlinear function $fal(\hat{\alpha}, \delta)$ given in [13, 14].

As shown in Fig. 3a, the PD controller exhibited a poor position control performance with a large steady state error between 1 and 3 seconds and a high sensitivity to external disturbances between 3 and 8 seconds (represented by the black curve in Fig. 3a). The steady-state error was eliminated by the SIMC, and the disturbance rejection characteristic improved as the free control parameter $\tau_c$ was decreased (depicted by the blue and green curves in Fig. 3a). However, this adjustment resulted in heightened noise sensitivity and a decrease in the stability of the robust motion controller. Notably, the SIMC-based controller exhibited instability for $\tau_c$ values below 0.002.

Disturbance rejection was improved using the SMC-based robust motion controllers, as illustrated in Fig. 3b. While increasing the SMC gain enhanced robustness against disturbances, it also exacerbated the chattering effect, leading to higher mechanical wear and increased energy consumption (represented by the black and blue curves in Fig. 3b). This issue can be mitigated by incorporating a DOb into the SMC-based robust motion controller, as depicted by the green curve in Fig. 3b. However, due to the discontinuous nature of the controller, residual chattering remains evident, as shown in the figure.

To eliminate chattering and ensure robustness, a PD controller with control gains $K_p = 2.5$ and $K_d = 0.25$ was employed in the outer-loop, while disturbances were suppressed using three different observer design methods – specifically DOb, LADRC and NADRC – in the inner-loop. In DOb and LADRC, the observer parameters were tuned through trial and error, balancing noise sensitivity and disturbance estimation performance. The bandwidths of the observers were set to 350 rad/s in both DOb and LADRC, providing faster dynamic response than the outer-loop controller, as recommended in [12–15]. This corresponds to $\ell_0 = 0.275$ in the conventional DOb-based robust motion controller presented in Section III. The nonlinear function $fal(\hat{\alpha}, \delta)$, proposed by Han, was tuned with $\hat{\alpha} = 0.001$ and $\delta = 0.95$ in the NADRC. Notably, greater effort was required to properly adjust the additional nonlinear terms in the NADRC, e.g., decreasing $\hat{\alpha}$ led to vibration, while increasing its value degraded the performance of disturbance estimation. Figure 3c illustrates that the observer-based robust motion controllers can achieve high-performance regulation control through the effective suppression of internal and external disturbances. This performance is, however, highly dependent on the precise tuning of key design parameters, such as the observer gain $\ell_0$ and the nominal inertia parameter $J_{mn}$ employed in the conventional DOb synthesis, as specified in Eq. (25).

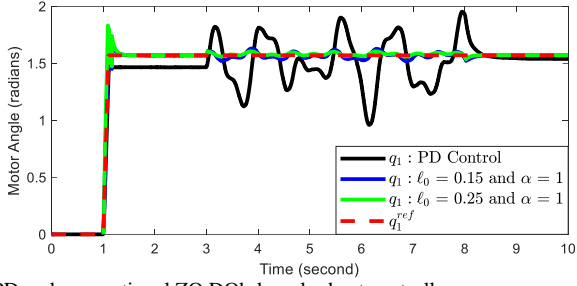

a) PD and conventional ZO DOb-based robust controllers.

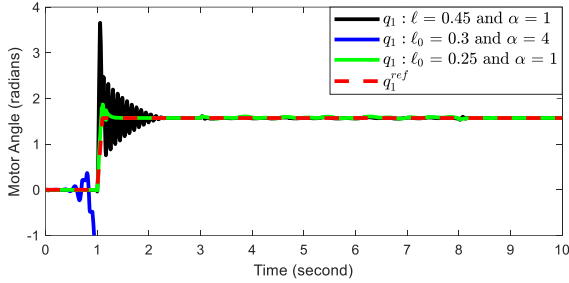

b) Stability of the conventional ZO DOb-based robust motion controller.

Fig. 4: Regulation control of Motor 1 using the PD controller with the control gains $K_p = 2.5$ and $K_d = 0.25$ and the ZO DOb-based digital robust motion controller. The same PD controller was employed in the outer-loop of the DOb-based robust motion controller.

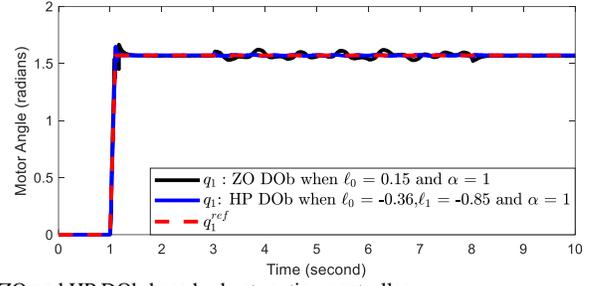

a) ZO and HP DOb-based robust motion controllers.

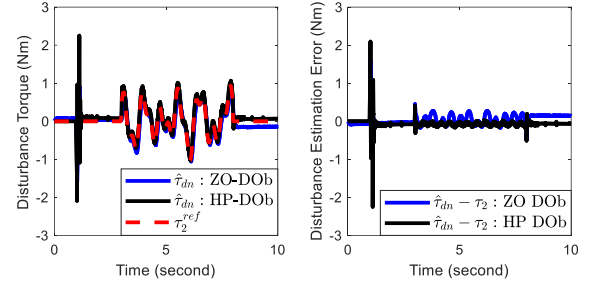

b) Estimated disturbance by the ZO and HP DObs.

Fig. 5: Regulation control of Motor 1 using the ZO and HP DOb-based digital robust motion controllers. The outer-loop PD controller gains were $K_p = 2.5$ and $K_d = 0.25$. $\tau_2$ and $\tau_2^{ref}$ represent the disturbance torque exerted by Motor 2 and its reference in regulation control, respectively.

### c) Stability and Performance of DOb-based Robust Motion Controller:

Figure 4 illustrates the influence of key design parameters on the stability and performance characteristics of the conventional ZO DOb-based robust motion controller. Proper tuning of these parameters significantly enhances position control performance through the implementation of the ZO DOb in the inner-loop. While the DOb effectively eliminates steady-state error, its external disturbance suppression capability can be further improved by increasing the observer gain, as shown by the blue and green curves in Figure 4a. Nevertheless, Fig. 4b shows that the observer gain cannot be freely increased due to the stability constraints derived in Sections III and IV. The DOb-based robust motion controller exhibited significant oscillations, indicating poor stability, when the observer gain was increased from $\ell_0 = 0.25$ to $\ell_0 = 0.45$ (as shown by the green and black curves in Fig. 4b). The stability constraint given in Eq. (24) explains this poor position control performance. Figure 4b also shows that not only the observer gain but also the nominal inertia parameter affects the stability of the digital robust motion controller. Although the observer gain was decreased to $\ell_0 = 0.3$, the position control system exhibited an unstable response when $\alpha = 4$, i.e., $J_{mn} = 4J_m$, as shown by the blue curve in Fig. 4b. To achieve robust stability and high-performance, the design parameters of the digital motion controller should therefore be tuned using the stability constraint in Eq. (39).

### d) HP DOb-based Robust Motion Controller:

Let us now demonstrate how the proposed HP DOb can enhance regulation control performance. Figure 5 presents the regulation control experiments conducted with both the ZO and HP DObs implemented in the inner-loop. The design parameters of both DObs were tuned to ensure that identical bandwidth values were utilised for disturbance estimation, allowing for a fair comparison of their performance in regulating the motion control system. As indicated by the blue and black curves in Fig. 5a, the HP DOb demonstrated superior robustness in suppressing disturbances compared to the ZO DOb, thereby enhancing the overall robust performance and stability of the system. The proposed HP DOb improves the robustness of the digital motion controller by providing more accurate disturbance estimation than the conventional DOb, as evidenced by the blue and black estimated disturbance curves and their respective errors shown in Fig. 5b. The robust motion controllers exhibited comparable levels of sensitivity to noise, while the disturbance estimation performance was significantly improved by the HP DOb. This superior disturbance estimation capability can lead to improved robust stability and high-performance in practical motion control systems.

### e) High-Order DOb-based Robust Motion Controller:

Similarly, Fig.6 illustrates the regulation control experiments conducted with both the ZO and FO DObs employed in the inner-loop when their bandwidths were set to the same value for a fair comparison. Figures 6a and 6b demonstrate that the FO DOb outperforms the conventional DOb, providing more accurate regulation control in the presence of disturbances. Moreover, it enables the estimation of both disturbances and their first-order derivatives, as illustrated in Figs. 6b and 6c, enhancing the system's ability to respond to dynamic changes in external inputs. However, compared to the ZO DOb, the FO DOb exhibited increased sensitivity to noise due to the use of the derivative of disturbances in the observer synthesis.

### f) Trajectory Tracking Control:

Lastly, let us present the trajectory tracking control experiment using the proposed digital robust motion controllers. In this experiment, Motor 1 was subjected to the same disturbances exerted by Motor 2 for 5 seconds while

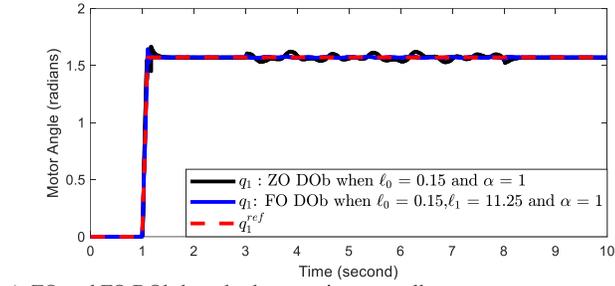

a) ZO and FO DOb-based robust motion controllers.

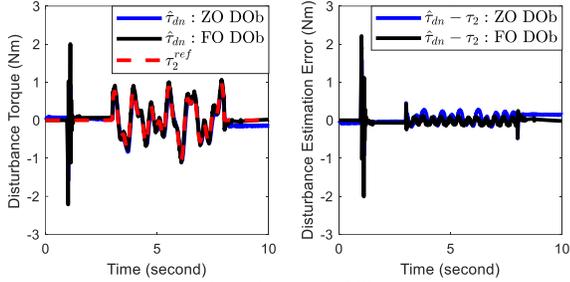

b) Estimated disturbance by the ZO and FO DOb.

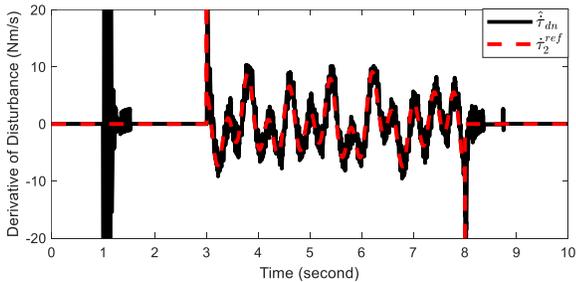

c) Derivative of the first order disturbance reference and its estimation.

Fig. 6: Regulation control of Motor 1 using the ZO and FO DOb-based digital robust motion controllers. The outer-loop PD controller gains were $K_p = 2.5$ and $K_d = 0.25$. $\dot{\tau}_2$ and $\dot{\tau}_2^{ref}$ represent the derivative of the disturbance torque exerted by Motor 2 and its reference in regulation control, respectively.

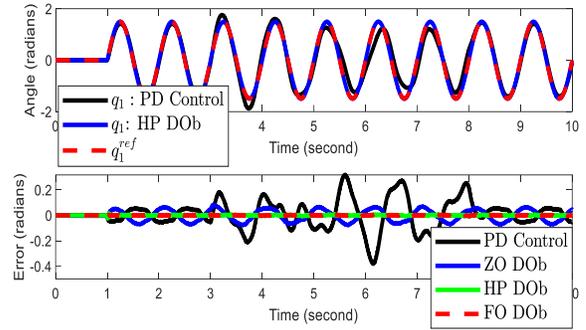

Fig. 7: Trajectory tracking control. The outer-loop PD controller gains were $K_p = 2.5$ and $K_d = 0.25$.

tracking a sinusoidal reference trajectory with a 1 Hz frequency between 1 and 10 seconds. For clarity, the trajectory tracking results are presented for the PD controller and the HP DOb-based robust motion controller only, while the error plots are illustrated for the PD controller as well as the ZO, FO and HP DOb-based robust motion controllers. The bandwidths of the DObs were set to the same value to ensure a fair comparison of their performance in disturbance estimation and robust trajectory tracking. Figure 7 demonstrates that the proposed robust motion controllers enable high-performance trajectory tracking by effectively suppressing both internal and external disturbances. Notably, the system exhibited superior performance, with smaller trajectory tracking errors when the HP and FO DObs were employed in the inner-loop, highlighting their enhanced ability to reject disturbances and improved tracking accuracy.

## VI. CONCLUSION

This paper has focused on the rigorous analysis and synthesis of DOb-based digital robust motion control systems in state-space, providing a comprehensive approach for enhanced performance and stability. It is demonstrated that the proposed synthesis method substantially streamlines the design process for advanced digital robust motion controllers. The method enables the development of an advanced DOb that improves disturbance estimation accuracy, while allowing for the synthesis of conventional ZO and HO DObs within the same systematic design framework. This approach can pave the way for high-performance robust motion control applications across various engineering fields. A novel high-performance DOb has been proposed by incorporating a more realistic discrete disturbance model into the DOb synthesis process. Further research is warranted to develop advanced DOb designs that achieve superior disturbance estimation accuracy and enhanced motion control performance. The rigorous stability analysis presented in the paper shows that the design parameters of the digital DObs should be properly tuned to achieve stability and high-performance. This paper analytically derives the stability constraints of the digital DObs using Lyapunov's direct method. The proposed design constraints serve as a critical tool for the systematic implementation of DOb-based digital robust motion control systems, mitigating stability and performance variations due to intuitive design approaches in practical applications. Future work should explore the extension of the proposed analysis and synthesis method to digital explicit and implicit robust force control systems. Additionally, it is essential to evaluate the application of digital robust motion controllers in advanced engineering systems, such as compliant and soft robots.

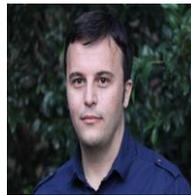

**Emre Sariyildiz** (S'11, M'16, SM'22) received his Ph.D. degrees in Integrated Design Engineering from Keio University, Tokyo, Japan, in September 2014, and in Control and Automation Engineering from Istanbul Technical University, Istanbul, Turkey, in February 2016. He was a research fellow in the Department of Biomedical Engineering and the Singapore Institute for Neurotechnology (SINAPSE) at the National University of Singapore, Singapore, from 2014 to 2017. Since April 2017, he has been with the School of Mechanical, Materials, Mechatronic, and Biomedical Engineering at the University of Wollongong, NSW, Australia, where he currently serves as a Senior Lecturer. His main research interests include control theory, robotics, mechatronics, and motion control.